\documentclass[12pt,preprint]{aastex}

\newcommand{\lsim}{\raise0.3ex\hbox{$<$}\kern-0.75em{\lower0.65ex\hbox{$\sim$}}}
\newcommand{\gsim}{\raise0.3ex\hbox{$>$}\kern-0.75em{\lower0.65ex\hbox{$\sim$}}}
\newcommand{\propsim}{\raise0.3ex\hbox{$\propto$}\kern-0.75em{\lower0.65ex\hbox{$\sim$}}}

\begin{document}
    
\title{Implications of the low frequency turn-over in the spectrum of radio knot C in DG Tau}

\author{C.-I. Bj\"ornsson\altaffilmark{1}}
\altaffiltext{1}{Department of Astronomy, AlbaNova University Center, Stockholm University, SE--106~91 Stockholm, Sweden.}
\email{bjornsson@astro.su.se}

\begin{abstract}
The synchrotron spectrum of radio knot C in the protostellar object DG Tau has a low frequency turn-over. This is used to show that its magnetic field strength is likely to be at least 10\,mG, which is roughly two orders of magnitude larger than previously estimated. The earlier, lower value is due to an overestimate of the emission volume together with an omission of the dependence of the minimum magnetic field on the synchrotron spectral index. Since the source is partially resolved, this implies a low volume filling factor for the synchrotron emission. It is argued that the high pressure needed to account for the observations is due to shocks. In addition, cooling of the thermal gas is probably necessary in order to further enhance the magnetic field strength as well as the density of relativistic electrons. It is suggested that the observed spectral index implies that the energy of the radio emitting electrons is below that needed to take part in first order Fermi acceleration. Hence, the radio emission gives insights to the properties of its pre-acceleration phase. Attention is also drawn to the similarities between the properties of radio knot C and the shock induced radio emission in supernovae. 
\end{abstract}


\section{Introduction}
Magnetic fields play an important role in many astrophysical phenomena. In contrast to density and temperature, which can often be reliably determined through spectral diagnostics, the strength of the magnetic field is more elusive. Since charged particles are accelerated by a magnetic field, observations of the resulting radiation gives information of its strength as well as direction. For non-relativistic electrons (cyclotron radiation), this gives directly the magnetic field strength. The draw-back is that this is a very weak emission process and, hence, hard to observe; in addition, it is often overwhelmed by other emission processes. Relativistic effects can increase the radiative efficiency dramatically. As a result, synchrotron emission from relativistic electrons is often observed in situations where particles can be accelerated to high energies. However, in contrast to cyclotron radiation, synchrotron radiation does not directly give the magnetic field strength, since the radiated frequency depends not only on the magnetic field strength but also the energy of the emitting particles. In order to get an estimate of the magnetic field strength, additional information is needed. The most commonly used effects are synchrotron self-absorption and inverse Compton scattering from the same electrons, which are responsible for the synchrotron radiation. However, both of these effects are model dependent; in particular, the result is quite sensitive to source inhomogeneities \citep{b/k17}. This is so even for partially resolved sources, since scales smaller than the spatial resolution cannot be probed.

In spite of the relativistic enhancement, synchrotron radiation is a rather feeble emission mechanism. This can be seen in extra-galactic observations, which, historically, have been dominated by two types of synchrotron sources. The low pressure regions of the large scale structures of radio galaxies is delineated by synchrotron radiation. In order for synchrotron radiation to become significant in a high pressure environment, an extra ingredient is normally needed; for example, bulk relativistic motion in the jets emanating from the central black hole in blazars. In the latter situation, the magnetic field is likely to play a role also for the launching as well as the collimation of the jet \citep{b/z77, b/p82}.

Jets seem to be the universal agent carrying away the angular momentum needed in order for accretion to proceed either onto a compact central object or to the formation of a star. In galactic sources, jet velocities are, with a few exceptions, non-relativistic. However, there are indications that a number of jet characteristics are rather insensitive to its velocity or energy carried by the outflow, e.g., the formation and collimation of the jet and, in particular, the role played by the magnetic field \citep{liv11}. In this respect, it is of interest that in the last years radio synchrotron emission has been detected from some of the least energetic jets associated with low-mass protostars \citep{r/f21}. The radio emission is observed in the outer part of the outflow, likely due to its interaction with the surrounding medium, rather than the inner region close to the protostar.

DG Tau is a nearby, well observed low-mass protostar with a prominent outflow, which is dominated by a narrow jet with blue-shifted outflow velocities of a few hundred km/s \citep{m/f83}. This jet is part of a bipolar outflow, where the radiation from the receding jet is substantially suppressed, most likely due to absorption from an accretion disk close to the protostar itself \citep{lav97}. Two regions of non-thermal emission have been detected, each of which associated with one of the components of the bipolar outflow. It now seems clear that neither of these synchrotron emitting regions is directly part of the narrow jets, since both of them lie some distant away from the deduced jet axes. It has been suggested that at least one of them is part of a bow shock from the approaching jet. This has been questioned by \cite{pur18}, since its proper motion is quite small, in fact, it is consistent with being stationary. Hence, the relation of the non-thermal emission to the narrow jet is still an open question.

It is thought that the narrow jets are just one part in a much broader outflow from the protostar. It is therefore possible, that the non-thermal emission regions may throw some light on this, otherwise, hard to observe region. However, as discussed above, in order to realistically constrain the properties of these regions, observations of the optically thin synchrotron spectrum is usually not enough. The advent of sensitive, low frequency telescopes like the Giant Metrewave Radio Telsecope (GMRT) \citep{ana05} and the  Low-Frequency Array (LOFAR) \citep{van13} opens up the possibility to search for a turn-over in the spectrum. The detection of such a low frequency turn-over would be important for an increased understanding of the source structure and, possibly, its formation.

\cite{fee19} established that a low-frequency turn-over is present in one, and possibly both, of the synchrotron emission regions in DG Tau. The implications for the strength of the magnetic field depend on the mechanism causing the turn-over. Synchrotron self-absorption is a possibility; however, as argued by \cite{fee19}, a more likely cause is the thermal plasma present in the outflow. In the latter case, either free-free absorption or the Razin effect is suggested.

The focus of this paper is how to best constrain the magnetic field from observations. Section \ref{sect2} discusses basic synchrotron theory. For optically thin emission, it is common to consider only the relativistic electrons responsible for the observed radiation. It is shown that starting, instead, from a given electron distribution, and then derive the radiation in the observed frequency band, can give magnetic field strengths differing by an order of magnitude. The reason is the neglect of the observed spectral index in the first method. Section \ref{sect3} is concerned with the effects a thermal plasma can have on the low frequency part of an otherwise optically thin synchrotron spectrum. It is shown that the ratio between the observed synchrotron and free-free fluxes is a convenient observable to use when constraining the magnetic field. When no free-free emission is observed, this provides a lower limit to the magnetic field. Section \ref{sect4} emphasizes that some rather model independent conclusions can be drawn from the radio observations of DG Tau; for example, that the strength of the magnetic field is likely to be at least two orders of magnitude larger than earlier estimated. In section \ref{sect5}, the radio observation are set into a broader context, where also optical observations are included. The relation between the narrow jet and the non-thermal emission is discussed in a few different shock scenarios. It is also argued that the radio observations indicate that the standard treatment of particle acceleration may need a revision. Lastly, a few useful future observations are suggested. The conclusions of the paper are collected in section \ref{sect6}. Numerical results are mostly given using cgs-units. When this is the case, the units are not written out explicitly.

\section{Determining the strength of the magnetic field in a synchrotron source} \label{sect2}
A homogeneous synchrotron source can be described by its emission volume ($V$), the magnetic field ($B$) and the energy distribution of relativistic electrons $n(\gamma)$. Normally, the latter is taken to be a power-law, i.e., $n(\gamma) = K_{\rm o} \gamma^{-{\rm p}}$ for $\gamma > \gamma_{\rm min}$ and $\rm {p\,>\,2}$, where $K_{\rm o}$ and $\gamma_{\rm min}$  are constants. In an optically thin source, the spectral flux is then given by $F_{\nu} \propto \nu^ {-(p-1)/2}$, where $\nu$ is the frequency. In some cases, the source is spatially resolved. Usually, this does not help to constrain $V$, since the degree of inhomogeneity is hard to establish. Hence, the two possible observables, i.e., $F_{\nu}$ and p, are not enough to deduce the source properties. Various assumptions are then needed in order to elucidate the characteristics of the source. One of the most popular is the assumption that the total energy content of the source is close to its minimum value consistent with the observed $F_{\nu}$.

\subsection{Minimum energy requirement}\label{sect2a}
The total energy of the source can be written $E_{\rm tot} = U_{\rm B} + U_{\rm rel}$, where $U_{\rm B} = VB^2/8\pi$ and $U_{\rm rel}$ is the total energy of relativistic particles. The latter is usually taken to be proportional to the total energy of electrons radiating at $\nu$, i.e., $U_{\rm rel} =a \gamma mc^2 N(\gamma)$ , where $a$ is a constant, $N(\gamma)$ is the total number of electrons with Lorentz factor $\gamma$ and $mc^2$ is the rest mass energy of the electron. With
\begin{equation}
	\nu F_{\nu} =\frac{\sigma_{\rm T}c\gamma^2 B^2N(\gamma)}{6\pi},
	\label{eq1}
\end{equation}
where $\sigma_{\rm T}$ is the Thomson cross-section \citep{r/l04}, one finds
\begin{equation}
	U_{\rm rel} = 6\pi\frac{amc}{\sigma_{\rm T}}\left(\frac{\nu_{\rm o}}{\nu}\right)^{1/2}\frac{\nu F_{\rm \nu}}{B^{3/2}}.
	\label{eq2}
\end{equation}
Here, $\nu = \gamma^2 \nu_{\rm B}$, where $\nu_{\rm B} = eB/2\pi mc\,(\equiv B \nu_{\rm o})$  is the cyclotron frequency. It should be noted that it is implicitly assumed that at a given frequency only one electron energy contributes to the emission (see also below). $E_{\rm tot}$ can then be minimised with respect to the value of $B$. The corresponding value of the magnetic field is denoted by $B_{\min}$.

However, in general, $a$ is not a constant but, for a given electron distribution, depends on $B$. As will now be shown, this omission can have a large effect on $B_{\min}$. In order to elucidate the $B$-dependence of $a$, it is instructive to derive an approximate expression along the traditional lines. For simplicity, let $U_{\rm rel}$ denote the total energy of relativistic electrons only and $U_{\rm rel}(>\gamma)$ the total energy of relativistic electrons with Lorentz factors larger than $\gamma$. They are related by
\begin{equation}
	U_{\rm rel}(>\gamma) = \left(\frac{\gamma}{\gamma_{\rm min}}\right)^{2-{\rm p}}U_{\rm rel}
	\label{eq3}
\end{equation}
With the use of $U_{\rm rel}(>\gamma)$ instead of $\gamma mc^2 N(\gamma)$ in equation (\ref{eq1}), equation (\ref{eq2}) is substituted by 
\begin{equation}
	U_{\rm rel} = 6\pi\frac{mc\nu_{\rm o}}{\sigma_{\rm T}\gamma_{\rm min}^{{\rm p}-2}}\left(\frac{\nu}{\nu_{\rm o}}\right)^{(\rm p -1)/	
	2}\frac{F_{\rm \nu}}{B^{(\rm p +1)/2}}.
	\label{eq4}
\end{equation}
Again, minimising $E_{\rm tot}$ with respect to $B$, one finds 
\begin{equation}
	B_{\rm min} = \left\{\frac{6\pi({\rm p+1}) e}{\sigma_{\rm T}\gamma_{\rm min}^{{\rm p}-2}}\left(\frac{\nu}{\nu_{\rm o}}\right)^{(\rm p -1)/	
	2}\frac{F_{\nu}}{V}\right\}^{2/({\rm p+5})}.
	\label{eq5}
\end{equation}

This shows the explicit p-dependence of $B_{\rm min}$. For p=2, this reduces to the standard expression often used, i.e., a= constant. The reason is that for p\,=\,2, the logarithmic distribution of electron energies is constant, i.e., the total energy of electrons radiating at a given frequency is $U_{\rm rel}/\ln(\gamma_{\rm max}/\gamma_{\rm min})$, which is independent of frequency. Here, $\gamma_{\rm max}$ is the upper cut-off in the electron energy distribution. Furthermore, since $U_{\rm rel}(>\gamma)$ includes the energy of all the electrons with a Lorentz larger than $\gamma$, it somewhat overestimates the energy of those electrons radiating at the frequency corresponding to $\gamma$. Hence, the value of $B_{\rm min}$ obtained from equation (\ref{eq5}) is a lower limit to the actual value (cf. equation \ref{eq8a}). One may also note that $\nu^{(\rm p-1)/2}F_{\nu}$ is independent of frequency.

In the general case, when the magnetic fields differs from $B_{\rm min}$, let $U_{\rm B}/U_{\rm rel} =\kappa$. In a similar manner, this leads to an approximate value for the magnetic field
\begin{equation}
	B_{\rm app} = \left(\frac{4\kappa}{{\rm p+1}}\right)^{2/({\rm p+5})} B_{\rm min},
	\label{eq7}
\end{equation}
so that $\kappa = {\rm (p+1)/4}$ for $B = B_{\rm min}$.
Hence, the corresponding value of the magnetic field can be expressed as
\begin{equation}
	B_{\rm app} = \left\{\frac{ 24\pi \kappa}{\gamma_{\rm min}^{{\rm p}-2}}\frac{e}{\sigma_{\rm T}}\left(\frac{\nu}{\nu_{\rm o}}\right)^{(\rm p -1)/	
	2}\frac{F_{\nu}}{V}\right\}^{2/({\rm p+5})}.
	\label{eq8}
\end{equation} 

However, in most cases, there is no reason to prefer either of the above approximate values of the magnetic field to the correct one, which can be obtained directly from the synchrotron emissivity \citep{r/l04, pac70},
\begin{equation}
	j_{\nu}^s = \frac{1}{16 \sqrt{\pi}}\frac{\sigma_{\rm T}}{e} ({\rm p-2})\gamma_{\rm min}^{\rm p-2} \chi({\rm p}) u_{\rm rel} B^{\rm (p+1)/2}
	\left(\frac{\nu_{o}}{\nu}\right)^{\rm (p-1)/2},
	\label{eq8a}
\end{equation}
where $u_{\rm rel} \equiv U_{\rm rel}/V$ and $\chi ({\rm p})$ is given in the Appendix. Since $j_{\nu}^s \equiv F_{\nu}/V$, one finds from equation (\ref{eq8a})
\begin{equation}
	B = \left\{\frac{128 \pi^{3/2} \kappa}{\chi({\rm p})({\rm p}-2)\gamma_{\rm min}^{{\rm p}-2}}\frac{e}{\sigma_{\rm T}}
	\left(\frac{\nu}{\nu_{\rm o}}\right)^{(\rm p -1)/2}\frac{F_{\nu}}{V}\right\}^{2/({\rm p+5})}.
	\label{eq9}
\end{equation}
The function $\chi({\rm p})$ varies slowly with ${\rm p}$; e.g., for an isotropic distribution of electrons $\chi({\rm 2})=5.38$ and $\chi({\rm 3})= 4.73$. For ${\rm p=2}$, $({\rm p-2})\gamma_{\rm min}^{\rm p-2}$ is replaced by $1/\ln(\gamma_{\rm max}/\gamma_{\rm min})$. It is seen that $B_{\rm app}$ is a rather good approximation for $B$; e.g.,  $B/B_{\rm app} = 1.19$ for ${\rm p=3}$. As mentioned above, the somewhat larger value for  $B$ as compared to $B_{\rm app}$ is due to the fact that the latter is calculated using an approximation, which overestimates the energy density of the radiating electrons. The effect of the observed spectral index ($\alpha \equiv({\rm p-1})/2$) on the deduced value for $B$ is directly obtained from equation (\ref{eq9}); for example,
\begin{equation}
	\frac{B({\rm p=3})}{B({\rm p=2})} = 5.22\left(\frac{V_{45}}{\kappa F_{\nu,16}}\right)^{1/28}\frac{\nu_{9}^{3/28}}{\gamma_{\rm min}^{1/4}
	\ln ^{2/7}(\gamma_{\rm max}/\gamma_{\rm min})}.
	\label{eq10}
\end{equation}
Although the source parameters have been scaled to be appropriate for DG Tau; i.e., $V_{45} \equiv V/10^{45}$, $F_{\nu,16} \equiv F_{\nu}/10^{16}$ and $\nu_{9} \equiv \nu /10^9$, it is seen that their actual values only marginally affect the sensitivity of the value of $B$ to the measured spectral index. In the case of DG Tau, \cite{fee19} find $\rm p \approx 4$, which increases the ratio in equation (\ref{eq10}) to, roughly, $10$. Hence, the deduced value of the magnetic field can vary by an order of magnitude depending on the observed spectral index. 

\subsection{Optically thick synchrotron radiation}\label{sect2b}
When the transition to the optically thick part of the spectrum is observed, the frequency where the spectral flux peaks ($\nu_{\rm abs}$) provides an additional constraint on the source properties. As is shown in the Appendix, the emission weighted average of the Lorentz factors ($\gamma_{\rm abs}$) of the electrons contributing to the radiation at $\nu_{\rm abs}$ is defined by $\gamma^2_{\rm abs}\equiv \nu_{\rm abs} / \nu_{\rm B}$ and its value can be obtained from
\begin{equation}
	\gamma_{\rm abs} =  \left[\frac{3\pi}{2^{11}}\frac{\eta^2({\rm p})\,y^2}{g({\rm p})\tau_{\rm abs}^2\{1-\exp(-\tau_{\rm abs})\}}\right]^{\rm 1/
	(2p+13)}\left(\frac{F_{\nu_{\rm abs}}}{mc^2}\right)^{\rm 1/(2p+13)}
	\label{eq11}
\end{equation}
where $\tau_{\rm abs}$ is the optical depth at $\nu_{\rm abs}$. The function $\eta({\rm p})$ is proportional to the absorption and $g({\rm p}) = \chi({\rm p})/\eta({\rm p})$. Furthermore, $y \equiv ({\rm p -2})\gamma_{\rm min}^{\rm p -2}(R_{||}/R)/\kappa$, where $R_{||}$ is the extension along the line of sight and $R$ the projected radius of the source, respectively (i.e., assuming the projection to be circular, gives a volume $V =\pi R^2R_{||}$, see Appendix). In the same way as above, for ${\rm p=2}$, $({\rm p-2})\gamma_{\rm min}^{\rm p-2}$ is replaced by $1/\ln(\gamma_{\rm max}/\gamma_{\rm min})$. One may note that, in general, the relation between $\nu$ and $\nu_{\rm B}$ is defined by $\gamma^2 \equiv \nu / \nu_{\rm B}$. As shown in the Appendix, $\gamma$ (and, hence, $\gamma_{\rm abs}$) is a slowly decreasing function of ${\rm p}$. This is due to the fact that, at a given frequency, the relative contribution of lower energy electrons to $F_{\nu}$ increases for a steeper spectrum. 
Furthermore, the definition of brightness temperature implies that its value at $\nu_{\rm abs}$ is given by
\begin{equation}
	T_{\rm b,abs} = \frac{g(p)\{1-\exp(-\tau_{\rm abs})\}}{3}\frac{\gamma_{\rm abs}mc^2}{k}.
	\label{eq12a}
\end{equation}
Hence, the radius of the source can be expressed as
\begin{equation}
	R = \frac{1}{2\pi}\left[\frac{3}{2g({\rm p})\{1-\exp(-\tau_{\rm abs})\}}\right]^{1/2}
	\left(\frac{F_{\nu_{\rm abs}}}{m\nu_{\rm abs}^2\gamma_{\rm abs}}\right)^{1/2}. 
	\label{eq12}
\end{equation}

With scalings appropriate for DG Tau and $\rm p=3$, equation (\ref{eq11}) yields
\begin{equation}
	\gamma_{\rm abs} = 1.6 \times 10 \left(y^2 F_{\nu_{\rm abs,16}} \right)^{1/19}.
	\label{eq13}
\end{equation}
\cite{fee19} deduce $\rm p \approx 4$ for DG Tau; however, as discussed below, this value is rather uncertain and, instead, $\rm p=3$ will be used when numerical values are given. The strength of the magnetic field is then obtained from $B = (\nu_{\rm abs}/\nu_{\rm o})/\gamma_{\rm abs}^2$.
Likewise, from equation (\ref{eq12}), the radius of the emitting surface is given by
\begin{equation}
	R = 3.4 \times 10^{11} \frac{F_{\nu_{\rm abs,16}}^{9/19}}{y^{1/19} \nu_{\rm abs,9}},
	\label{eq14}
\end{equation}
where, $\tau_{\rm abs}(\rm p=3) = 0.64$ has been used (see Appendix). 

An important property of synchrotron radiation should be noted; namely, while $R$ depends quite sensitively on both $F_{\nu_{\rm abs}}$ and $\nu_{\rm abs}$, the value of $\gamma_{\rm abs}$ is independent of $\nu_{\rm abs}$ and depends weakly on $F_{\nu_{\rm abs}}$. In fact, it is the latter property that lies behind the p-dependence of $B$ in equation (\ref{eq9}): The brightness temperature at $\nu_{\rm abs}$ is proportional to $\gamma_{\rm abs}$ (equation \ref{eq12a}). Hence, for a given value of the brightness temperature at optically thin frequencies $T_{\rm b} \propto \gamma_{\rm abs} (\nu_{\rm abs}/\nu)^{(\rm p+3)/2}$. Since $\gamma_{\rm abs}$ is independent of $\nu_{\rm abs}$, this shows directly that a larger value for p corresponds to a larger value for $\nu_{\rm abs}$. Furthermore, the insensitivity of the value of $\gamma_{\rm abs}$ to  $F_{\nu_{\rm abs}}$ implies that the increased value of $\nu_{\rm abs}$ is due mainly to a higher value for $B$.

An alternative derivation of the expression for $B$ in equation (\ref{eq9}) can be given, starting with the properties of the brightness temperature at $\nu_{\rm abs}$. At optically thin frequencies, $F_{\nu} \propto R^2 \nu^2 T_{\rm b} \propto R^2 \nu^2 \gamma_{\rm abs} (\nu_{\rm abs}/\nu)^{(\rm p+3)/2}$ or $\nu^{(\rm p-1)/2} F_{\nu} \propto R^2 \gamma_{\rm abs}^{\rm p+4} B^{(\rm p+3)/2}$. It is shown in the Appendix that $\gamma_{\rm abs}$ can also be expressed as (cf. equation \ref{eq11}) 
\begin{equation}
	\gamma_{\rm abs} =  \left\{\frac{3\pi}{2^{11}}\frac{\eta^2({\rm p})\,y^2}{g({\rm p})\tau_{\rm abs}^3}
	\left(\frac{\nu}{\nu_{\rm B}}\right)^{\rm (p-1)/2} \frac{F_{\nu}}{mc^2}\right\}^{\rm 1/3(p+4)}.
	\label{eq15}
\end{equation}
This implies $\nu^{(\rm p-1)/2} F_{\nu} \propto R^3 y B^{(\rm 5+p)/2}$ or
\begin{equation}
	B \propto \left\{\frac{\kappa}{({\rm p-2})\gamma_{\rm min}^{\rm p-2}} \frac{\nu^{(\rm p-1)/2} F_{\nu}}{V}\right\}^{2/(\rm 5+p)},
	\label{eq16}
\end{equation}
where $\pi R^2 R_{||} = V$ has been used.

\section{Implications of a low frequency turn-over in a synchrotron spectrum}\label{sect3}
It was shown in the previous section that observation of the synchrotron self-absorption frequency provides an estimate of the emitting surface (equation \ref{eq14}). In addition to synchrotron self-absorption, the most likely mechanisms to cause a low frequency turn-over are free-free absorption and the Razin effect \citep{fee19}. The aim in this section is to deduce constraints on the size of the synchrotron source that can be obtained from either of these mechanisms. Together with equation (\ref{eq9}), the result can then be used to find the strength of the magnetic field. Both of these mechanisms involve a thermal plasma. Hence, it proves convenient in the analysis to scale the observed optically thin synchrotron emission to the free-free emission, which must be present at some level.

\subsection{The Razin effect}\label{sect3a}
Synchrotron emission can be seen as a relativistic effect, which lengthens the phase coherence of the cyclotron emission process. The Razin effect results from an increased phase velocity of light, due to the presence of a thermal plasma, which reduces the relativistic enhancement at low frequencies. Since the thermal plasma must coexist with the synchrotron emitting electrons, one may write $F_{\nu}/V = j^{\rm s}_{\nu}\equiv x j^{\rm ff}_{\nu}$, where $ j^{\rm ff}_{\nu}$ is the free-free spectral emissivity per unit volume. The constant of proportionality ($x$) is an observable. However, even if free-free emission would be observed together with the synchrotron emission, the deduced value of $x$ should be regarded as a lower limit, since free-free emission from outside the synchrotron source may contribute to the observed value. 

The free-free emissivity is given by $ j^{\rm ff}_{\nu} = 6.8\times10^{-38}n_{\rm e}^2/T^{\rm 1/2}$ \citep{tuc75}, where $n_{\rm e}$ is the density and $T$ the temperature of the thermal electrons in the synchrotron source. The connection between the synchrotron emission and the thermal plasma is expressed through the Razin frequency $\nu_{\rm R}= 20\,n_{\rm e} /B$. With the use of $x$ and $\nu_{\rm R}$, the spectral emissivity of the synchrotron radiation can be written
\begin{equation}
	\frac{F_{\nu}}{V} = 1.7\times10^{\rm -23}\frac{B^2 \nu_{\rm R,9}^2 x_{\rm1}}{T_{\rm 4}^{\rm 1/2}}.
	\label{eq17}
\end{equation}
Here, the Razin frequency is normalized as $\nu_{\rm R,9} = \nu_{\rm R}/10^9$, $T_{4} =T/10^4$ and $x_{1} = x/10$. For convenience, it has been assumed that the plasma consists of protons and electrons only and, furthermore, the Gaunt factor has been set equal to unity.

The adopted spectral index of the synchrotron emission in DG Tau has varied between different authors; for example, \cite{ain14} find $\alpha = 0.89$, while \cite{fee19} give $\alpha \approx 1.5$, depending on how the spectral fitting is made. This difference may be due to the short spectral range over which the emission is observed and, also, possible time variations can have an effect, since the flux measurements at the different frequencies are not all coeval. As already mentioned, all numerical results for DG Tau will be given assuming $\alpha = 1$, i.e., $p=3$.

When the expression for $F_{\nu}/V$ in equation (\ref{eq17}) is substituted into equation (\ref{eq9}), one finds
\begin{equation}
	B_{\rm R} = 2.6\times10^{-2}\left(\frac{\kappa}{\gamma_{\rm min}}\right)^{1/2}\frac{(x_{\rm R,1}\nu_{9})^{1/2} \nu_{\rm R,9}}
	{T_{\rm R,4}^{1/4}},
	\label{eq18}
\end{equation}
where the subscript $"R"$ indicates that the values apply for the case when the spectral turn-over is due to the Razin effect. Furthermore, equations (\ref{eq17}) and (\ref{eq18}) can be combined to find the corresponding emission volume
\begin{equation}
	V_{\rm R, 45} = 8.8\times 10^{-4}\,\frac{\gamma_{\rm min}}{\kappa}\frac{T_{\rm R,4}}{\nu_{\rm R,9}^4}\frac{\nu_{9}F_{\nu, 16}}
	{(\nu_{9} x_{\rm R,1})^2}.
	\label{eq19}
\end{equation}
Note that $\nu_{9}F_{\nu, 16}$ as well as $\nu_{9} x_{1}$ are independent of frequency.

\subsection{Free-free absorption}\label{sect3b}
When the low frequency turn-over is due to free-free absorption, the comparison between the synchrotron and free-free emission is most conveniently done with the use of the corresponding brightness temperatures. The ratio between the synchrotron and free-free emissions can then be written 
\begin{equation}
	x_{\rm ff} = \left(\frac{\nu_{\rm abs}}{\nu_{\rm ff}}\right)^{\rm (p+3)/2} \frac{T_{\rm b,abs}^{\rm thin}}{T_{\rm ff}},
	\label{eq20}
\end{equation}
where $T_{\rm b,abs}^{\rm thin}$ is the optically thin brightness temperature of the synchrotron emission extrapolated to $\nu_{\rm abs}$ (see Appendix). Furthermore, $T_{\rm ff}$ is the temperature of the thermal plasma in the free-free absorption region. In contrast to the Razin effect, the ratio between the synchrotron and free-free emission ($x_{\rm ff}$) needs to be measured at $\nu_{\rm ff}$, where $\nu_{\rm ff}$ is the frequency where the optical depth to free-free absorption is unity. The value of $ x_{\rm ff}$ corresponds to the free-free emission, which covers the synchrotron source. Just as in the Razin case, a measured value for $ x_{\rm ff}$ should be regarded as a lower limit, since the free-free emission region may be more extended than the synchrotron source. Furthermore, the synchrotron flux measured at $\nu_{\rm ff}$ is that extrapolated from the optically thin part of the spectrum (i.e., somewhat higher than the actual value at $ \nu_{\rm ff}$).
The value of $B$ can be deduced directly from equation (\ref{eq20}) as
\begin{equation}
	B_{\rm ff} = \frac{\nu_{\rm ff}}{\gamma_{\rm abs}^2 \nu_{o}} \left(\frac{x_{\rm ff}\,T_{\rm ff}}{T_{\rm b,abs}^{\rm thin}}\right)^{\rm 2/(p+3)},
	\label{eq21}
\end{equation}
where the subscript "ff" indicates that this value of $B$ applies when the spectral turn-over is due to free-free absorption. 

With the use of $T_{\rm b,abs}^{\rm thin} = 6.3\times10^8\gamma_{\rm abs}$ and
\begin{equation}
	\gamma_{\rm abs}= 1.6\times 10\left\{\left(\frac{R_{\rm ||}}{R}\frac{\gamma_{\rm min}}{\kappa}\right)^2 \frac{\nu_{9} F_{\nu,16}}{B}
	\right\}^{1/21}
	\label{eq22}
\end{equation}
one finds
\begin{equation}
	B_{\rm ff} = 2.0\times 10^{-2}\nu_{\rm ff,9}^{9/8}\left(\frac{\kappa}{\gamma_{\rm min}}\frac{R}{R_{\rm ||}}\right)^{1/4}
	\left(\frac{x_{\rm ff,1}^3\,T_{\rm ff,4}^3}{\nu_{9} F_{\nu,16}}\right)^{1/8},
	\label{eq23}
\end{equation}
where $\nu_{\rm ff,9} = \nu_{\rm ff}/10^9$. The corresponding emission volume is obtained from equation (\ref{eq9}) as
\begin{equation}
	V_{\rm ff, 45} = 2.3\times 10^{-3}\frac{R_{||}}{R\,\nu_{\rm ff,9}^{9/2}}\left(\frac{\nu_{9}F_{\nu,16}}{x_{\rm ff,1}\,T_{\rm ff,4}}\right)^{3/2}
	\label{eq24}
\end{equation}

\subsection{A comparison between the Razin effect and free-free absorption}\label{sect3c}
With the use of $V = \pi R^2 R_{||}$ in equation (\ref{eq24}), one finds
\begin{equation}
	R_{\rm ff,14} = 9.1\times 10^{-1}\frac{1}{\nu_{\rm ff,9}^{3/2}}\left(\frac{\nu_{9}F_{\nu,16}}{x_{\rm ff,1}T_{\rm ff,4}}\right)^{1/2}.
	\label{eq25}
\end{equation}
This is the analog of equation (\ref{eq12}) for the synchrotron self-absorption case. It is seen that free-free absorption gives no constraint on the extension of the source along the line of sight, only its projected size. The reason is that a turn-over due to absorption can only give information on the emitting surface. This differs from the Razin case, which causes a turn-over in the optically thin emission and, instead, gives an expression for the total emission volume (eq. \ref{eq19}). The implications of this difference is discussed further below. 

It is of interest to compare the values of $B_{\rm R}$ and $B_{\rm ff}$ deduced from a given observed turn-over frequency $\nu_{\rm obs}$. Although a detailed spectral fitting is likely to give somewhat different values for $\nu_{\rm R}$ and $\nu_{\rm ff}$, the approximations $\nu_{\rm R}\approx \nu_{\rm ff} \approx \nu_{\rm obs}$ in the expressions above should be appropriate. In order to facilitate further, it is convenient to measure $x$ and $F_{\nu}$ at $\nu = \nu_{\rm obs}$. It is then found from equations (\ref{eq18}) and (\ref{eq23}) that
\begin{equation}
	\frac{B_{\rm ff}}{B_{\rm R}} = 7.8\times 10^{-1}\left\{\frac{\gamma_{\rm min}R}{\kappa R_{||}}\right\}^{1/4} \frac{T_{\rm ff,4}^{3/8} 
	T_{\rm R,4}^{1/4}}{\nu_{\rm obs,9}^{1/2} x_{\rm obs,1}^{1/8} F_{\nu_{\rm obs},16}^{1/8}}, 
	\label{eq26}
\end{equation}
where $\nu_{\rm obs,9} = \nu_{\rm obs}/10^9$.

It is sometimes argued \citep{fee19} that assuming the low frequency turn-over to be due to the Razin effect would give a lower limit to the value of the magnetic field. However, it is seen from equation (\ref{eq26}) that this may not generally be the case. This can be illustrated by the situation when the Razin effect and the free-free absorption are both caused by the same thermal plasma. Since $\nu_{\rm ff} = 0.13 n_{\rm e}R_{||}^{1/2}/T^{3/4}$ and $n_{\rm e} = \nu_{\rm R} B/20$, one finds
\begin{equation}
	\frac{\nu_{\rm ff}}{\nu_{\rm R}} = 6.7\times 10\frac{BR_{||,14}^{1/2}}{T_{4}^{3/4}}
	\label{eq27}
\end{equation}
It is seen from equations (\ref{eq26}) and  (\ref{eq27}) that it is possible to have, for example, $\nu_{\rm ff} > \nu_{\rm R}$ together with $B_{\rm ff} < B_{\rm R}$.

The thermal plasma, which give rise to the Razin  effect, will also cause free-free absorption. In general then, the ratio in equation (\ref{eq27}) is a lower limit to the actual value, i.e., including also absorption from an external plasma. If it can be established (e.g., through spectral fitting) that a low frequency turn-over is due to the Razin effect, i.e., $\nu_{\rm R} > \nu_{\rm ff}$, equation (\ref{eq27}) implies a lower limit to the temperature, which can be used in equation (\ref{eq18}) to obtain 
\begin{equation}
 	B_{\rm R} < 2.4\times 10^{-2}\left(\frac{\kappa}{\gamma_{\rm min}}\right)^{3/8}\frac{\left(x_{\rm R,1} \nu_{9}\right)^{3/8}\nu_{\rm R,9}
	^{3/4}}{R_{||,14}^{1/8}}.
	\label{eq28}
\end{equation}
The corresponding lower limit to the temperature is
\begin{equation}
	T_{\rm R,4} > 1.8 \left(\frac{\kappa}{\gamma_{\rm min}}\right)^{1/2}\left(x_{\rm R,1} \nu_{9}\right)^{1/2}\nu_{\rm R,9} R_{||,14}^{1/2}.
	\label{eq29}
\end{equation}

\subsection{The connection between the thermal and non-thermal properties of the plasma}\label{sect3d}
It is common to parameterize the energy densities in relativistic particles and magnetic field in terms of the thermal energy density, $u_{\rm th} = 3kTn_{\rm e}$, so that $u_{\rm rel} \equiv \kappa_{\rm e} u_{\rm th}$ and $u_{\rm B}\equiv \kappa_{\rm B} u_{\rm th}$ (hence, $\kappa = \kappa_{\rm B}/\kappa_{\rm e}$). The implied values for $ \kappa_{\rm e}$ and $ \kappa_{\rm B}$ can then be used to constrain the physical mechanisms responsible for accelerating the electrons/protons and enhancing the magnetic field strength; for example, when both of these aspects of the plasma are related to the presence of a shock, one expects the values of both $ \kappa_{\rm e}$ and $ \kappa_{\rm B}$ to be less than unity, unless cooling of the thermal gas is important (see below). 

For the Razin case, $F_{\nu} \propto j^{\rm ff}_{\nu} \propto n_{\rm e}^2 \propto B^2$. Since $u_{\rm B} = B^2/8\pi$, it is seen from equation (\ref{eq9}) that the $B$-dependence cancels for $p=3$ and an expression for $u_{\rm rel}$ can be obtained directly. From equation (\ref{eq18}) one finds
\begin{equation}
	u_{\rm rel} = 2.7\times 10^{-5}\frac{\nu_{\rm R,9}^2 x_{R,1}\nu_{9}}{\gamma_{\rm min}T_{\rm R,4}^{1/2}}.
	\label{eq30}
\end{equation}
Instead, an expression for $B$ can be found from $u_{\rm B} = \kappa_{\rm B}u_{\rm th} = 3\kappa_{\rm B}kT\nu_{\rm R}B/20$, where $n_{\rm e} = \nu_{\rm R}B/20$ has been used. This yields
\begin{equation}
	B_{\rm R} = 5.2\times 10^{-3} \kappa_{\rm B}T_{\rm R,4}\nu_{\rm R,9}.
	\label{eq31}
\end{equation}
Furthermore, using $u_{\rm B}/u_{\rm rel} = \kappa_{\rm B}/\kappa_{\rm e}$, one finds
\begin{equation}
	\kappa_{\rm B}\kappa_{\rm e} = 2.5\times 10 \frac{x_{R,1}\nu_{9}}{\gamma_{\min} T_{\rm R,4}^{5/2}}.
	\label{eq32}
\end{equation}
It is seen that the $\nu_{\rm R}$-dependence disappears; hence, this expression relates the non-thermal and thermal properties of any plasma independent of whether or not a low frequency turn-over is observed.

In general, the synchrotron spectral emissivity depends on $p$. Hence, one would expect an explicit $B$-dependence in equation (\ref{eq32}). However, for the same reason as the cancellation of $B$ in equation (\ref{eq9}), for $p=3$ the relation between the thermal and non-thermal properties of the plasma is expressed in a particular simple form. This can be seen by rewriting equation (\ref{eq8a}) as
\begin{equation}
	j_{\nu}^s = \frac{\chi({\rm p})({\rm p-2})\gamma_{\rm min}}{8\sqrt{\pi}}\frac{\sigma_{\rm T}}{e}\frac{\nu_{\rm o}u_{\rm rel}u_{\rm B}}
	{\nu}\left(\frac{\gamma_{\rm min}}{\gamma}\right)^{\rm p-3}
	\label{eq33}
\end{equation}
This shows that for $p=3$, the synchrotron emissivity is $ \propto u_{\rm rel}u_{\rm B}\gamma_{\rm min}/\nu$. Since the free-free emissivity can be expressed as $j^{\rm ff}_{\nu} \propto u_{\rm th}^2/T^{5/2}$, one finds that $x = j^{\rm s}_{\nu}/j^{\rm ff}_{\nu}$ leads to $\kappa_{\rm B}\kappa_{\rm e} \propto  x\nu/(\gamma_{\rm min}T^{5/2})$.

When the low frequency turn-over is due to free-free absorption, the expression for the $B$-field in equation (\ref{eq23}) results in
\begin{equation}
	u_{\rm B}u_{\rm rel} = 2.7\times 10^{-10}\nu_{\rm ff,9}^{9/2}\frac{R}{R_{||}\gamma_{\rm min}}\frac{(x_{\rm ff,1}T_{\rm ff,4})^{3/2}}
	{(\nu_{9}F_{\nu,16})^{1/2}}.
	\label{eq34}
\end{equation}
The temperature in equation (\ref{eq34}) is that pertaining to the absorption region, which may be external to the synchrotron source. Hence, there are no direct constraints on the thermal energy density in the synchrotron emission region. Although comparing equation (\ref{eq34}) to equation (\ref{eq30}) may suggest that it is roughly of the same order of magnitude as when the low frequency turn-over is due to the Razin effect, one should note the different temperature dependences in the Razin and free-free cases.

\section{Observations}\label{sect4} 
The results in sections 2 and 3 assume a power-law distribution of relativistic electrons. However, as already mentioned, the determination of the spectral index of the optically thin emission is affected by a fair amount of uncertainty and $\alpha = 1\,({\rm i.e.,}\,p=3 )$ was taken as a representative value. This introduces some uncertainty in the values of the observables deduced from the observations. It was decided to use 1\,GHz as reference frequency and scale values to other frequencies using $p=3$.

\cite{fee19} judged the Razin effect to give a better spectral fit to the data than free-free absorption. However, data are such that free-free absorption cannot be excluded, in particular, when a slightly inhomogeneous external source is allowed for. Since no value was given for a possible free-free absorption frequency, for convenience, the same value as for the Razin frequency will be used for $\nu_{\rm ff}$ below, i.e., $\nu_{\rm R} = \nu_{\rm ff} = 6.3\times10^8$. With the use of the GAIA-distance of 121 pc \citep{bai18}, the observed flux density of 0.50 mJy at 1\,GHz gives $\nu_{9}F_{\nu,16} = 0.88$. Furthermore, with a flat free-free emission spectrum and no clear indication of a flattening at the highest observed frequencies, $\nu_{9}x_{1} \gsim 1$. 

For a low frequency turn-over due to the Razin effect, equations (\ref{eq18}) and (\ref{eq19}) then imply $B_{\rm R}\,\gsim1.6\times 10^{-2} (\kappa/\gamma_{\rm min})^{1/2}/ T_{\rm R,4}^{1/4}$ and $V_{\rm R,45}\,\lsim\,4.9\times 10^{-3}(\gamma_{\rm min} T_{\rm R,4})/\kappa$.
Assuming instead the low frequency turn-over to be caused by free-free absorption, one may deduce from equations (\ref{eq23}) and (\ref{eq25}) that $B_{\rm ff}\,\gsim\,1.2\times 10^{-2}(\kappa R/\gamma_{\rm min} R_{||})^{1/4}\,T_{\rm ff,4}^{3/8}$ and $R_{\rm ff,14}\,\lsim\,1.7\,T_{\rm ff,4}^{-1/2}$.

Although synchrotron self-absorption is an unlikely cause for the low frequency turn-over, the implied values for the source size and magnetic field are useful as references when considering the implications of the values derived above assuming a thermal origin. With $B = \nu_{\rm abs}/(\nu_{\rm o}\gamma_{\rm abs}^2)$ and $\nu_{\rm abs} = 6.3\times 10^8$, one finds from equations (\ref{eq13}) and (\ref{eq14}) that $B=0.93\,y^{-4/19}$ and $R= 5.0 \times 10^{11} y^{-1/19}$, where $y =\gamma_{\rm min}R_{||}/(\kappa R)$.

A few rather model independent conclusions can be drawn from the above results; a more detailed discussion is given in the next section. The deduced source properties depend on the unknown temperature in the emission ($T_{\rm R}$)/absorption ($T_{\rm ff}$) region. For the range of temperatures expected in the jet environment of DG Tau, it is seen that the corresponding values of $B_{\rm R}$ and $B_{\rm ff}$ are rather similar. Furthermore, the temperature dependence is not strong enough to avoid the conclusion that the strength of the magnetic field is likely to be at least two orders of magnitude larger than normally deduced. Since only a lower limit to $x$ is obtained from observations, it may be noted that an actual value of $x_{1} \sim 10^2 -10^3$ would lead to a magnetic field strength of the same order of magnitude as its synchrotron value.

Knot C is spatially resolved \citep{pur18} and implies a source size a few times $10^{15}$. In contrast to the strength of the magnetic field deduced from observation, the choice between a low frequency turn-over due to the Razin effect and free-free absorption has important implications for the source geometry. In the free-free absorption case, the observed size is at least an order of magnitude larger than the synchrotron emitting region. This argues for a low covering factor, i.e., no larger than $10^{-2}$. The Razin effect indicates that the synchrotron emission volume cannot be much larger than $10^{43}$; for a spherical source, the observations then suggest a low volume filling factor, i.e., no larger than $10^{-3}$. Alternatively, in this case, the source may be homogeneous so that the emission volume is a thin sheet with an extension along the line of sight of less than $10^{12}$; i.e., $R_{||}/R\,\lsim\,10^{-3}$. 

\section{Discussion}\label{sect5}
For physical conditions expected to prevail in the environment of the jet in DG Tau, a lower limit to the strength of the magnetic field is $\sim 10$\,mG. This value implies a thermal pressure corresponding to $n_{\rm e}T \sim 10^{10}/\kappa_{\rm B}$. In situations where the strength of the magnetic field and the acceleration of particles are due to processes behind a shock, $\kappa_{\rm B}\,<\,1$ is expected. When cooling is important, the energy density of the magnetic field may dominate the local thermal pressure. However, its value is unlikely to exceed the thermal energy density just behind the shock. Hence, independent of the importance of cooling for the synchrotron emission, the thermal energy density behind the shock is such that $n_{\rm e}T \sim 10^{10}$ should be a lower limit. 

This minimum value is more than three orders of magnitude larger than the value used in \cite{fee19}, which, in turn, was based on an analysis of emission line spectra taken by \cite{oh15}. However, the slit used by \cite{oh15} was aligned with the jet axis and had a width of 2.9 arc-second. Since the synchrotron emission from radio knot C lies some distance away from the jet axis, it is not clear that the thermal properties of radio knot C can be taken to be similar to those along the jet axis. In fact, the observations indicate that they are not. In the literature, the name "knot C" is sometimes used for both the optical emission region and the radio emission region. To avoid confusion, in the following, these emission regions will be denoted by "knot C" and "radio knot C", respectively. 

\cite{oh15} show the variations of emission line profiles along the jet axis. Although they are rather similar for most of the jet, knot C stands out. Here, most of the emission lines are much broader due to a low velocity extension; in particular, [SII]$\lambda6731$ is dominated by this low velocity component. Since the emission line ratio [SII]$\lambda$6731/[SII]$\lambda$6716 is a good density indicator \citep{ost74}, this suggests that the low velocity gas has a density in excess of $10^4$. 

\cite{rod12} observed DG Tau in a filter including [SII]$\lambda$6731 as well as [SII]$\lambda$6716. At the position of knot C, there is a diffuse extension from the jet axis in the direction of the radio knot C. It is possible that this extension corresponds to the low velocity component seen in the spectrum of knot C. Unfortunately, neither velocity nor density information can be obtained from these observations. It may be noted that the extent of the [SII] protrusion in the direction of radio knot C is large enough that it is likely that not all of its emission was covered by the slit used by \cite{oh15}.

\subsection{Shock scenarios}\label{sect5a}
The high pressure in radio knot C suggests that its emission comes from a region behind a shock. In a one dimensional flow, a constant input of momentum gives rise to a two component structure bounded by a reverse and a forward shock. Let the momentum input be due to a flow with velocity $v_{\rm jet}$ and density $n_{\rm jet}$, which impacts on a stationary external medium with density $n_{\rm s}$. Neglecting velocity differences in between the shocks, one finds $v_{\rm jet} +3v_{\rm r} = 3v_{\rm s}$, where $v_{\rm r}$ and $v_{\rm s}$ are the velocities of the reverse and forward shocks, respectively. Conservation of momentum then gives $n_{\rm jet}(v_{\rm jet}-v_{\rm r})^2 = n_{\rm s}v_{\rm s}^2$. This leads to $v_{\rm r} =( v_{\rm jet}/3)[4/\{1+(n_{\rm s}/n_{\rm jet})^{1/2}\}-1]$ and $v_{\rm s} =( 4v_{\rm jet}/3)/\{1+(n_{\rm s}/n_{\rm jet})^{1/2}\}$.

In DG Tau, the momentum input is likely coming from the outflow associated with the jet and the external medium corresponds to the surrounding molecular cloud. Hence, $n_{\rm s}>n_{\rm jet}$ is expected. One may note that for $n_{\rm s}>9\,n_{\rm jet}$, $v_{\rm r}$ becomes negative, i.e., the reverse shock moves upstream. In a jet geometry, instead of an upstream motion of the reverse shock, there will be a sideways outflow from the shocked region. The velocity of this outflow is roughly the sound speed. When cooling is not important, this gives rise to a back-flow, which envelopes the jet in a cocoon. However, as discussed below, cooling is expected to be important in DG Tau and, hence, the outflow velocity will be reduced. Rather than a cocoon, the outflow may then resemble a bow shock.

With the conditions prevailing in DG Tau, one expects $v_{\rm r} \ll v_{\rm jet}$. Since the observed velocities in the jet are a few hundred km/s, this implies a temperature $T \approx 10^6$ behind the reverse shock and, from the above discussion, $n_{\rm e}\,\gsim\,10^4$. The temperature behind the forward shock would be factor $n_{\rm s}/n_{\rm jet}$ smaller. If the synchrotron emission comes from the region behind the reverse shock, cooling of the thermal gas is likely to enhance the energy densities of the magnetic field as well as that of the relativistic electrons. Hence, the emission region is expected to lie some distance away from the reverse shock. For a magnetic field direction parallel to the shock front, $ B\propto n_{\rm e}$, so that $u_{\rm B}\propto n_{\rm e}^2$. In the same way, $u_{\rm rel} \propto n_{\rm e}^{4/3}$. Since the direction of the magnetic field can have a component perpendicular to the shock front, the ratio $u_{\rm B}/u_{\rm rel}$ may not vary too much during the cooling phase.

The pressure is expected to stay roughly constant during the cooling phase. With a starting temperature of $T\,\sim\,10^6$, cooling can increase the density by two orders of magnitude. The values of $u_{\rm B}$ and $u_{\rm rel}$ increase until equipartition is reach between the thermal and non-thermal components of the plasma. Although the density stays constant beyond this point, cooling can continue and, hence, the synchrotron emission region may have $\kappa_{\rm B}\,\gsim\,1$ and $\kappa_{\rm e}\,\gsim1$, even though their values where substantially below unity close to the shock front (as expected in shock acceleration scenarios). Note, also, that $\gamma_{\rm min} \propto n_{\rm e}^{1/3}$ during the cooling phase. 

If the synchrotron emission would come from the region behind the forward shock, cooling should be less important. The values of $\kappa_{\rm B}$ and $\kappa_{\rm e}$ are then expected to be quite small. Hence, the value of $\gamma_{\rm min}$ needs to be quite large in order for equation (\ref{eq32}) to be satisfied. At the same time, it cannot be larger than the value corresponding to the observed synchrotron emission. With $B\,\gsim\,10$\,mG, this shows that $\gamma_{\rm min}$ cannot be much larger than $10^2$. However, such a value is not consistent with the small values of $\kappa_{\rm B}$ and $\kappa_{\rm e}$ expected in a shock acceleration scenario. This argues in favour of the synchrotron emission coming from behind the reverse shock.

The unknown relation between knot C and radio knot C gives rise to two plausible overall scenarios for the the synchrotron emission. If radio knot C is directly related to knot C, the radio emission is likely coming from a region down-stream of the outflow from a small reverse shock (Mach disk) centred on the jet axis. On the other hand, it is possible that the reverse shock is due to a much wider outflow surrounding the observed narrow jet. In this case, there is not necessarily a direct connection between radio knot C and knot C. In turn, this could lead to a much broader reverse shock and, thus, the outflow from the shocked region would be less prominent. The geometry of the emission region in these two scenarios is likely to be quite different. As discussed in section\,\ref{sect4}, the possible geometries allowed by observations are constrained by the origin of the low energy cut-off, i.e., whether it is due to the Razin effect or free-free absorption. This issue together with a more detailed discussion of the emission structure behind a shock will be given in a forthcoming paper. 

\subsection{Particle acceleration}\label{sect5b}
First order Fermi-acceleration at shocks is thought to be the mechanism producing the relativistic electrons giving rise to the observed synchrotron emission in many astrophysical sources. Some properties of this scenario are well established; for example, that strong non-relativistic shocks result in power-law distributions of protons with p=2 and that electrons can start to take part in first order Fermi-acceleration when their Lamor radius is of the same order as that of the thermal protons (Bohm diffusion). However, several aspects are still not well understood; for example, the initial phase when not only the electrons but also the protons are pre-accelerated before entering first order Fermi-acceleration. The details of this phase determine the efficiency of injection of particles from the thermal pool into the acceleration process and, in particular, how this differs between protons and electrons. In the last years, detailed particle-in-cell (PIC) calculations have been able to simulate the formation of shocks from first principles, in which acceleration of particles as well as magnification of the magnetic field strength are integral parts \citep{c/s14a, c/s14b}. It should be noted that PIC-simulations are rather limited in scope, in that they only reach a small distance down-stream of the shock. This means also that the acceleration process can be followed only up to rather modest energies. However, this limitation should still allow a realistic treatment of the important phase in which particles are injected and pre-accelerated before entering first order Fermi-acceleration.

The PIC-simulations presented in \cite{par15} show the acceleration of electrons to be similar to that of protons, although the momentum/Lamor radius needed for injection into first order Fermi-acceleration is somewhat larger than for protons. For the low shock velocities expected in radio knot C, the energy of the relativistic electrons radiating in the radio regime is well above the limit indicated by the PIC-simulations, where first order Fermi acceleration should apply. Although the radio spectral index in radio knot C is somewhat uncertain, it is clear that p is substantially larger than 2. This does not accord with the PIC-simulations. It should be noted though, that the shock velocity in DG Tau is much smaller than that used in \cite{par15}. Hence, it is possible that an extrapolation of the results to the low velocities in DG Tau is not allowed. However, in this context, it is interesting to compare with radio supernovae, for which the spectral indices imply $p\approx3$ \citep{c/f06}. Here, shock velocities are comparable to those used in \cite{par15}. Furthermore, the shock velocities and magnetic field strengths are high enough so that, with the assumptions made in \cite{par15}, the observed radio emission is due to electrons with Lorentz factors smaller than the first order Fermi-acceleration limit and, hence, belong to the pre-acceleration phase. However, this is at odds with the results in \cite{par15}, which show  p $\approx$ 2 also in the pre-acceleration phase.  

Before dismissing first order Fermi-acceleration in favour of, for example, acceleration due to turbulence, it should be noted that the characteristics of the pre-acceleration phase in PIC-simulations are sensitive to a few assumptions that need to be made in order to make calculations feasible. 1) An artificially high electron mass is used. This is shown to affect the injection of particles; in particular the ratio between electrons and protons. 2) The flow is assumed to be homogeneous. There are indications that an inhomogeneous flow may substantially affect the shock formation process \citep{g/f21}. Hence, one may argue that the discrepancy between observations and theory is due to the treatment of the pre-acceleration phase; for example, a larger momentum for the electrons to enter first order Fermi-acceleration together with pre-acceleration due to processes giving p $\approx$ 3 would be consistent with the observations of both radio knot C and radio supernovae. However, in such a scenario, the actual first order Fermi-acceleration is not directly observed and becomes apparent only at higher frequencies.

The ratio between the energy densities in relativistic protons and relativistic electrons is usually argued to be quite large \citep{b/k15}; i.e., $a\gg1$ in section \ref{sect2a}. When this is the case, $\kappa$ should be replaced by $a\kappa$ in the above formulae. It is seen that this would substantially increase the lower limit of the magnetic field as well as straining a realistic physical connection between the non-thermal and thermal components of the plasma. The conclusion that $a\gg1$ is usually based on the assumption that $p=2$ for the electrons is valid in their whole energy range, i.e., down to $\gamma_{\rm min}$. This is also the reason that PIC-simulations yield a large value for $a$ \citep{par15}, since $p\approx$ 2 in the pre-acceleration phase. For a given observed synchrotron flux, $u_{\rm rel} \propto (\gamma_{\rm obs}/\gamma_{\rm min})^{\rm {p-2}}$, where $\gamma_{\rm obs}$ is the Lorentz factor of the electrons producing the observed radiation. Hence, an extended pre-acceleration phase characterized by $p\approx 3$ together with $\gamma_{\rm min} \approx 1$ would increase the energy density of relativistic electrons so that $a \approx 1$ may apply. 

Another similarity between radio knot C and radio supernovae is that the volume filling factor of the synchrotron emission is likely to be quite small. In many radio supernovae, the X-ray emission is most directly explained as inverse Compton scattering of the optical emission. \cite{bjo13} compared the radio and X-ray fluxes from such supernovae and found that the observations could be most directly accounted for by a X-ray emission volume roughly a hundred times larger than that corresponding to the synchrotron emission. This can result from relativistic electrons filling up most of the region behind the shock, while the magnetic field structure is filamentary/turbulent with a small filling factor.

\subsection{Useful future observations}\label{sect5c}
In order to increase the understanding of the properties of radio knot C, the most important observable to constrain further is the ratio between synchrotron and free-free emission (i.e., $x$). This affects not only the value of the magnetic field but also the relation between the non-thermal and thermal properties of the plasma (i.e., equation \ref{eq32}). The latter property gives clues to the mechanisms, which determine the magnetic field strength and accelerates the relativistic electrons.

The value of $x$ should be most easily determined by observations at the highest possible frequencies, since an observed flattening of the spectrum could indicate a transition from synchrotron to free-free emission. As discussed in section \ref{sect3a}, the radio spectral index is somewhat uncertain. In part, this is due to the flux measurements at the highest frequencies ($\sim 10$\,GHz), which seem to differ, roughly, by a factor of 2 \citep{fee19}. Although the reason for this is unclear, one may note that the simultaneous observations by \cite{pur18} at  $6$\,GHz and $10$\,GHz indicate a substantial flatter spectrum than observed at lower frequencies. A confirmation of such a flattening would be crucial, since this would constrain the actual value of $x$.

As discussed in section \ref{sect5}, the [SII]-extension from knot C in the direction of radio knot C observed by \cite{rod12} could be the link needed to determine the relation between the two emission regions. Spectral information would make it possible to establish whether this extension is the origin of the low velocity and high density component indicated in the spectrum taken by \cite{oh15}. If so, this would lend support to the scenario wherein the reverse shock corresponds to a Mach disk centred on the jet axis and that the radio emission is produced in the cooling outflow from the shocked region. 

\cite{pur18} has argued that due to its small proper motion, radio knot C is unlikely to be part of a bow shock structure that also includes the fast moving knot C. However, the high pressure in the synchrotron emission region suggests that cooling of the thermal gas is important. The dynamics of the outflow from the region in between the forward and reverse shocks is sensitive to the cooling as well as the shock geometry. Detailed observations in line with those started by \cite{pur18}, may be used to disentangle the kinematics within the emission region and, hence, provide a means to elucidate the relationship between knot C and radio knot C.

\section{Conclusions}\label{sect6}

The main result of the present paper is that the strength of the magnetic field in radio knot C is likely to be significantly larger than normally deduced from observations. The earlier, low values are due to a combination of two effects; namely, an omission of the dependence of its minimum value on the synchrotron spectral index and that the emission region is smaller than previously estimated. Each of these effects increases the magnetic field by an order of magnitude. A stronger magnetic field has several consequences:

1) Since the source is partially resolved, the small emission region implies a low volume filling factor for the synchrotron emission or, alternatively, a special geometry.

2) The high pressure indicated by the magnetic field suggests that the radio emission comes from a region behind a shock.

3) The properties of the jet environment in DG Tau make it likely that cooling of the thermal gas is needed in order to increase the magnetic field as well as the density of relativistic electrons. 

4) If first order Fermi acceleration is at play, the observed spectral index suggests that the synchrotron emitting electrons are still in the pre-acceleration phase. Due to the likely model dependence of this phase, it also constrains the mechanisms responsible for the injection of particles.

5) The value of the magnetic field deduced in this paper is a lower limit. A few observations are suggested, which may help constrain its actual value and, at the same time, elucidate the structure and kinematics within the synchrotron emission region.

\newpage

\appendix

\begin{center}
{\bf Appendix}
\end{center}

\section{Basic synchrotron theory}

The synchrotron spectral emissivity per unity volume and solid angle is $j^s_{\nu,{\Omega}} = j^s_{\nu}/4\pi$. For an isotropic distribution of electron energies, this yields \citep{r/l04, pac70, g/s65}
\begin{equation}
	j^{\rm s}_{\nu,{\Omega}} = \frac{1}{64 \pi^{3/2}}\frac{\sigma_{\rm T}}{e} ({\rm p-2})\gamma_{\rm min}^{\rm p-2} \chi({\rm p}) u_{\rm rel} B^{\rm  
	(p+1)/2}\left(\frac{\nu_{o}}{\nu}\right)^{\rm (p-1)/2},
	\label{eqA1}
\end{equation}
where $K_{\rm o} = ({\rm p-2})\gamma_{\rm min}^{\rm p-2}u_{\rm rel}/mc^2$ has been used. Here,
\begin{equation}
	\chi({\rm p}) = \frac{3^{\rm (p+2)/2}}{\rm p+1}\frac{\Gamma\left(\frac{\rm 3p+19}{12}\right)\Gamma\left(\frac{\rm 3p-1}{12}\right)
	\Gamma\left(\frac{\rm p+5}{4}\right)}{\Gamma\left(\frac{\rm p+7}{4}\right)},
	\label{eqA2}
\end{equation}
where, $\Gamma(z)$ is the gamma function.

 The corresponding absorption coefficient is given by \citep{r/l04, pac70, g/s65}
 \begin{equation}
 \mu^{\rm s}_{\nu} = \frac{\pi^{3/2}}{4}\eta({\rm p})\frac{e}{mc^2}({\rm p-2})\gamma_{\rm min}^{\rm p-2}u_{\rm rel}B^{(p+2)/	2}
 	\left(\frac{\nu_{\rm o}}{\nu}\right)^{\rm (p+4)/2},
	\label{eqA3}
\end{equation}
where
\begin{equation}
	\eta({\rm p	}) = 3^{\rm (p+1)/2}\frac{\Gamma\left(\frac{\rm 3p+22}{12}\right)\Gamma\left(\frac{\rm 3p+2}{12}\right)
	\Gamma\left(\frac{\rm p+6}{4}\right)}{\Gamma\left(\frac{\rm p+8}{4}\right)}.
	\label{eqA4}
\end{equation}
The source function is defined by $S_{\nu}\,\equiv\,j^s_{\nu,{\Omega}}/\mu^{\rm s}_{\nu}$. This leads to
\begin{equation}
	S_{\nu} = \frac{2\,g({\rm p})}{3} m\nu^2 \left(\frac{\nu}{\nu_{\rm B}}\right)^{1/2},
	\label{eqA5}
\end{equation}
where
\begin{equation}
	g({\rm p}) =\frac{\chi({\rm p})}{\eta({\rm p})}.
	\label{eqA6}
\end{equation}

\subsection{Observed properties of a synchrotron source}
A model is needed to deduce source properties from observations. Consider a spherical shell with outer radius $R$ and thickness $\Delta R \ll R$. Assume that this volume is homogeneously filled with a synchrotron emitting plasma. Furthermore, it will be assumed that interior to $R-\Delta R$, the source is optically thick to the synchrotron emission coming from the back side of the source. The emitting volume is then $V = 2\pi R^2 \Delta R$. The average extension of the source along the line of sight is, roughly, $R_{||} = 2\Delta R$; hence, $V =\pi R^2 R_{||}$ so that this geometry applies also to a disk with radius $R$ and thickness $R_{||}$.

The optical depth at a given frequency is then $\tau_{\nu} = \mu_{\nu}^{\rm s} R_{||}$, which can be rewritten as
\begin{equation}
	\tau_{\nu} = \frac{\sqrt{\pi}}{2^5}\eta({\rm p}) \frac{eB}{mc^2} \left(\frac{\nu_{\rm B}}{\nu}\right)^{\rm (p+4)/2} yR,
	\label{eqA7}
\end{equation}
where 
\begin{equation}
	y \equiv \frac{(\rm {p-2})\gamma_{\rm min}^{\rm p-2}}{\kappa} \frac{R_{||}}{R}.
	\label{eqA8}
\end{equation}
The emitted flux is given by $F_{\nu} = 4\pi^2R^2 S_{\nu} \{1-\exp(-\tau_{\nu})\}$, or
\begin{equation}
	F_{\nu} = \frac{8\pi^2}{3}g({\rm p})m\nu^2\left(\frac{\nu}{\nu_{\rm B}}\right)^{1/2}R^2\{1-\exp(-\tau_{\nu})\}.
	\label{eqA9}
\end{equation}
With the use of equation (\ref{eqA9}), equation(\ref{eqA7}) can be rewritten as
\begin{equation}
 	\tau_{\nu} = \left[\frac{3\pi}{2^{11}}\frac{\eta^2({\rm p})\,y^2}{g({\rm p})\{1-\exp(-\tau_{\nu})\}}\right]^{1/2} \left(\frac{\nu_{\rm B}}{\nu}
	\right)^{\rm (2p+13)/4} \left(\frac{F_{\nu}}{mc^2}\right)^{1/2}.
	\label{eqA10}
\end{equation}

The synchrotron self-absorption frequency ($\nu_{\rm abs}$) is normally defined as the frequency where the spectral flux peaks. Since $\nu \propto \tau^{\rm -2/(p+4)}$, the corresponding optical depth ($\tau_{\rm abs}$) is obtained from $\exp (\tau_{\rm abs}) = 1 + (p+4)\tau_{\rm abs}/5$. Equations (\ref{eqA9}) and (\ref{eqA10}) can then be used to relate observed quantities to source properties
\begin{equation}
	\gamma_{\rm abs} =  \left[\frac{3\pi}{2^{11}}\frac{\eta^2({\rm p})\,y^2}{g({\rm p})\tau_{\rm abs}^2\{1-\exp(-\tau_{\rm abs})\}}\right]^{\rm 1/
	(2p+13)}\left(\frac{F_{\nu_{\rm abs}}}{mc^2}\right)^{\rm 1/(2p+13)}
	 \label{eqA11}
\end{equation}
and
\begin{equation}
	R = \frac{1}{2\pi}\left[\frac{3}{2g({\rm p})\{1-\exp(-\tau_{\rm abs})\}}\right]^{1/2}
	\left(\frac{F_{\nu_{\rm abs}}}{m\nu_{\rm abs}^2\gamma_{\rm abs}}\right)^{1/2},
	\label{eqA12}
\end{equation}
where $\gamma_{\rm abs} \equiv \sqrt{\nu_{\rm abs}/\nu_{\rm B}}$ is the emission weighted average Lorentz factor of the electrons contributing to the spectral flux at $\nu_{\rm abs}$. Furthermore, since the brightness temperature is defined by $T_{\rm b}\,\equiv\,(c^2/2k\nu^2)S_{\nu}\{1-\exp(-\tau_{\nu})\}$, its value at $\nu_{\rm abs}$ is given by
\begin{equation}
	T_{\rm b,abs} = \frac{g({\rm p})\{1-\exp(-\tau_{\rm abs})\}}{3}\frac{\gamma_{\rm abs}mc^2}{k}.
	\label{eqA13}
\end{equation}

In the optically thin part of the spectrum the value of $\nu^{\rm(p-1)/2} F_{\nu}$ is constant. Denote the extrapolated value of $F_{\nu}$ at 
$\nu_{\rm abs}$ by $F_{\nu_{\rm abs}}^{\rm thin}$. This leads to $F_{\nu_{\rm abs}}^{\rm thin} = F_{\nu_{\rm abs}}\tau_{\rm abs}/ \{1-\exp(-\tau_{\rm abs})\}$. This relation can be used in equation (\ref{eqA11}) to obtain an expression for $\gamma_{\rm abs}$ in situations when no self-absorption frequency is observed,
\begin{equation}
	\gamma_{\rm abs} =  \left\{\frac{3\pi}{2^{11}}\frac{\eta^2({\rm p})\,y^2}{g({\rm p})\tau_{\rm abs}^3}
	\left(\frac{\nu}{\nu_{\rm B}}\right)^{\rm (p-1)/2} \frac{F_{\nu}}{mc^2}\right\}^{\rm 1/3(p+4)}.
	\label{eqA14}
\end{equation}
Since $\nu_{\rm abs}$ is not observed, there is an explicit $B$-dependence in this expression. Likewise, extrapolating the optically thin brightness temperature to $\nu_{\rm abs}$ gives $T_{\rm b,abs}^{\rm thin} = T_{\rm b,abs} \tau_{\rm abs}/\{1-\exp(-\tau_{\rm abs})\}$.


\clearpage

\begin{thebibliography}{}
    
    \bibitem[Ainsworth et al.(2014)]{ain14} Ainsworth, R.E., Scaife, A.M.M., Ray, T.P., Taylor, A.M., Green, D.A., \& Buckle, J.V.,
    2014, \apj, 792, L18
    
    \bibitem[Ananthakrishnan(2005)]{ana05} Ananthakrishnan, S., 2005, in International Cosmic Ray Conference, 
    eds. Acharya, B.S., et al., Vol 10, 125
    
    \bibitem[Bailer-Jones(2018)]{bai18} Bailer-Jones, C.A.L., Rybizki, J., Fouesneau, M., Mantelet, G., \& Andrae, R., 2018,
    \aj, 156, 58
    
    \bibitem[Beck \& Krause(2015)]{b/k15} Beck. R., \& Krause. M., 2015, AN, 326, 414
    
    \bibitem[Bj\"{o}rnsson(2013)]{bjo13} Bj\"{o}rnsson, C.-I., 2013, \apj, 769, 65
    
    \bibitem[Bj\"{o}rnsson \& Keshavarzi(2017)]{b/k17} Bj\"{o}rnsson, C.-I., \& Keshavarzi, S.T., 2017, \apj, 841, 12
    
    \bibitem[Blandford \& Payne(1982)]{b/p82} Blandford, R.D., \& Payne, D.G., 1982, \mnras,
   199, 883
   
   \bibitem[Blandford \& Znajek(1977)]{b/z77} Blandford, R.D., \& Znajek, R.L., 1977, \mnras,
   179, 433
    
    \bibitem[Caprioli \& Spitkovsky(2014a)]{c/s14a} Caprioli, D., \& Spitkovsky, A., 2014a, \apj, 783, 91
    
    \bibitem[Caprioli \& Spitkovsky(2014b)]{c/s14b} Caprioli, D., \& Spitkovsky, A., 2014b, \apj, 794, 46
    
    \bibitem[Chevalier \& Fransson(2006)]{c/f06} Chevalier, R.A., \& Fransson, C., 2006, \apj, 651, 381
    
    \bibitem[Feeney-Johansson et al.(2019)]{fee19} Feeney-Johansson, A., Purser, S.J.D., Ray, T.P., et al., 2019, \apj, 885, L7
    
    \bibitem[Ginzburg \& Syrovatskii(1965)]{g/s65} Ginzburg, V.L., \& Syrovatskii, S.I., 1965, \araa, 3, 297
    
    \bibitem[Grassi \& Fiuza(2021)]{g/f21} Grassi. A., \& Fiuza. F., 2021, arXiv:2105.11750
    
    \bibitem[Lavalley et al.(1997)]{lav97} Lavalley, C., Cabrit, S., Dougados, C., Ferruit, P., \& Bacon, R., 1997,
    \aap, 327, 671
    
    \bibitem[Livio(2011)]{liv11} Livio, M., 2011, in " Gamma Ray Bursts", eds. McEnery, J.E., Racusin, J.L., \& Gehrels, N., 
    AIP Conf. Proc. 1358, 329
    
    \bibitem[Mundt \& Fried(1983)]{m/f83} Mundt, R., \& Fried, J.W., 1983, \apj, 274, L83
    
    \bibitem[Oh et al.(2015)]{oh15} Oh, H., Pyo, T.-S., Yuk, I.-S., \& Park, B.-G., 2015, JKAS, 48, 113
    
    \bibitem[Osterbrock(1974)]{ost74} Osterbrock, D.E., 1974, in "Astrophysics of Gaseous Nebulae", San Francisco, Freeman
    
    \bibitem[Pacholczyk(1970)]{pac70} Pacholczyk, A., 1970, in "Radio Astrophysics", San Francisco, Freeman
    
    \bibitem[Park et al.(2015)]{par15} Park, J., Caprioli, D., \& Spitkovsky, A., 2015, \prl, 114, 085003
    
    \bibitem[Purser et al.(2018)]{pur18} Purser, S.J.D., Ainsworth, R.E., Ray, T.P., Green, D.A., Taylor, A.M., \& Scaife, A.M.M.,  
    2018, \mnras, 481, 5532
    
    \bibitem[Ray \& Ferreira(2021)]{r/f21} Ray, T.P., \& Ferreira, J., 2021, NewAR, 93, 101615
    
    \bibitem[Rodr\'{i}guez et al.(2012)]{rod12} Rodr\'{i}guez, L.F., Gonz\'{a}lez, R.F., Raga, A.C., et al., 2012, \aap, 537, A123
    
    \bibitem[Rybicki \& Lightman(2004)]{r/l04} Rybicki, G.B., \& Lightman, A.P., 2004, in "Radiative Processes in Astrophysics", 
    Wiley-VCH, Weinheim
    
    \bibitem[Tucker(1975)]{tuc75} Tucker, W.H., 1975, in "Radiation Processes in Astrophysics", Cambridge, MIT Press
    
    \bibitem[Van Haarlem et al.(2013)] {van13} Van Haarlem, M.P., Wise, M.W., Gunst. A.W., et al., 2013, \aap, 556, A2
    
    
       
\end{thebibliography}
\end{document}